\documentclass{article}
\usepackage{spconf} 
\usepackage{amsmath}
\usepackage{graphicx}
\usepackage{subfig}
\usepackage[bookmarks=true,bookmarksnumbered=true,bookmarksopen=false,%
 breaklinks=true,pdfborder={0 0 0},backref=false,colorlinks=true,citecolor=blue]{hyperref}

\newcommand{\webpageurl}{https://www.speech.kth.se/tts-demos/}
\newcommand{\webpageurltext}{www.speech.kth.se/tts-demos/}

\title{Prosody-controllable spontaneous TTS with neural HMMs}
\name{Harm Lameris, Shivam Mehta, Gustav Eje Henter, Joakim Gustafson, Éva Székely\thanks{This research was supported by the Swedish Research Council projects Connected (VR-2019-05003), Perception of speaker stance 
(VR-2020-02396), the Riksbankens Jubileumsfond project CAPTivating 
(P20-0298), and by the Wallenberg AI, Autonomous Systems and Software Program (WASP) funded by the Knut and Alice Wallenberg Foundation.}}
\address{Division of Speech, Music and Hearing, KTH Royal Institute of Technology, Stockholm, Sweden}
\begin{document}
%
\maketitle
\begin{abstract}
Spontaneous speech has many affective and pragmatic functions that are interesting and challenging to model in TTS (text-to-speech).
However, the presence of reduced articulation, fillers, repetitions, and other disfluencies mean that text and acoustics are less well aligned than in read speech.
This is problematic for attention-based TTS.
We propose a TTS architecture that is particularly suited for rapidly learning to speak from irregular and small datasets while also reproducing the diversity of expressive phenomena present in spontaneous speech.
Specifically, we modify an existing neural HMM-based TTS system, which is capable of stable, monotonic alignments for spontaneous speech, and add utterance-level prosody control, so that the system can represent the wide range of natural variability in a spontaneous speech corpus.
We objectively evaluate control accuracy and perform a subjective listening test to compare to a system without prosody control.
To exemplify the power of combining mid-level prosody control and ecologically valid data for reproducing intricate spontaneous speech phenomena, we evaluate the system's capability of synthesizing two types of creaky phonation.
\end{abstract}


\section{Introduction}
In recent years, the quality of end-to-end deep neural-network-based Text-to-Speech (TTS) architectures such as Tacotron 2 \cite{tacotron} have improved to rival human speech \cite{an2021}. Two issues faced by state-of-the-art TTS are a lack of ecological validity \cite{wester2016} and expressivity in the learned prosody. Most architectures are trained on read-speech corpora, e.g., \cite{tacotron, glowtts} that have limited prosodic coverage, and generate ``average'' prosody, treating naturally occurring variation as random noise \cite{raitio2}.  Moreover, the prosody is generated solely from text, which allows no control over the generated style \cite{raitio2021}. Concurrently, spontaneous speech is increasingly used in TTS \cite{personality, adigwe22_interspeech}. Spontaneous speech data is challenging to model, due to disfluencies and large variability \cite{spontaneous}; offers high ecological validity for evermore commonplace conversational AI systems; and the varied prosody offered by spontaneous data positively impacts factors such as word recall and attention \cite{pitch}. For conversational systems it would also be useful to be able to synthesize creaky phonation, as this have been found to be a strong turn-yielding cue \cite{ogden2001turn, gravano2011turn}.  

Several approaches exist for prosody-controlled TTS. In \cite{reference_encoder}, the authors use a Tacotron 2 architecture for which the decoder is conditioned on a prosodic reference encoder. During training, the prosodic reference encoder generates prosody embeddings from  spectrogram slices in an unsupervised manner. At inference, the prosody of a reference audio file is transferred to the target audio. In \cite{vqvae} a quantized fine-grained VAE with an autoregressive prosody prior is used which learns a latent representation of the prosody from the aligned spectrogram.

The approaches closest to this paper appear in \cite{raitio2021} and \cite{shechtman2019sequence}. In \cite{raitio2021} the authors introduce a hierarchical model based on Tacotron 2 that uses a separate prosody encoder to predict sentence-wise pitch, phone duration, speech energy, and spectral tilts. This allows for the control of these features and the production of a variety of styles, while achieving similar mean opinion scores to a baseline Tacotron 2 model. This hierarchical model requires a large amount of data (36 hours of read-speech, 16 of conversational), for which sufficient-length spontaneous speech corpora do not exist, and requires enormous amount of resources to train (300k steps per model on 16 GPUs at a batch size of 512). In \cite{shechtman2019sequence} Tacotron 2 is modified by appending embedding values for the pitch, loudness, and duration to the output of the encoder before being passed to an augmented attention mechanism. Although the augmented attention mechanism increased robustness, the prosodic features require a large encoder, and combined with the large amount of training time required for attention mechanisms, this method is infeasible for spontaneous TTS. 

In this paper we study the effect of using spontaneous speech data for TTS with prosodic feature control:
We use a method based on neural hidden Markov models (HMM) TTS \cite{mehta2022neural}, equivalent to a kind of transducer TTS \cite{yasuda2019initial}. Important for spontaneous speech applications, neural HMMs \cite{tran2016unsupervised} force monotonic alignments between input symbols and output frames, which helps to train rapidly and on smaller, more disorderly datasets than neural TTS based on conventional neural attention \cite{mehta2022neural}.
The statewise nature of the neural HMM is also appealing for modelling disfluencies and other speech irregularities that have been transcribed with discrete tokens. 
Here, there is a possibility to represent speech phenomena such as disfluencies,
partial repetitions, under-articulated speech segments, 
etc.,\ that cause the alignment between speech and transcription to be lower in a spontaneous speech corpus than in a read speech corpus \cite{szekely2020augmented}.
We extend neural HMM TTS to learn a control space for the mean and variation of the fundamental frequency (f0) and speech rate (syllables per second).

As we use spontaneous data, and with the control of the f0, we are able to implicitly control the voice quality. More specifically, we can influence the presence or absence of creaky voice produced by the speech synthesizer. Vocal fry, discrete glottal pulses of low frequency \cite{fry}, has not been synthesized using neural TTS, as it is complex to measure due to its periodicity. By opting for synthesis with a low mean f0 and low f0 variation. We are able to synthesize both stylistic vocal fry and vocal fry that occurs at end-of-turn.

We perform a data analysis, an objective evaluation of the control of the feature modification space, an expert analysis of the creaky voice quality produced by the system, and a subjective evaluation of the synthesized speech of our system. Audio samples are provided at \href{\webpageurl}{\webpageurltext}.
\section{Method}
\subsection{Data}
We used two datasets for our speech synthesizers. For our base-model we used a scripted conversational corpus, RyanSpeech corpus \cite{ryanspeech}. This corpus contains 10 hours or 11,279 utterances  of a single male speaker of US English reading textual materials from real-world conversational settings, including chat and task-oriented dialogues designed for the Ryan Social Robot, and a random selection from the Libri-TTS corpus. The spontaneous model was trained on a corpus created from the audio of the Trinity Speech-Gesture dataset \cite{tsgd}, which consists of 25 impromptu monologues by a male voice actor speaking Hiberno-English. In the monologues, the voice actor speaks in a natural conversational style, in a colloquial manner. The actor spoke without interruptions from another speaker, and was instructed to speak about topics of his choosing. The monologues concern topics such as hobbies, daily activities, and interests.

We pre-processed the spontaneous corpus by segmenting the monologues into breath groups, i.e., single stretches of speech between two breath events, as was performed in \cite{personality, fps}. We opted for breath groups as a unit since we hypothesize that minimal style changes occur within a single breath group. Using breath groups also enables the possibility to change style within a given utterance by inserting a breath. The breath groups were combined into overlapping breath group bigrams to create audio files of up to 11 seconds \cite{szekely2020breathing}. We extracted the mean and standard deviation of the fundamental frequency and mean speech rate per breath group using the Wavelet Prosody Toolkit (WPT) \cite{wavelet} to create three prosodic features: f0 variability (the per utterance standard deviation of f0), pitch (mean f0), and speech rate (syllables per second). Although these features were chosen as the prosodic features for the experiment, other prosodic features such as energy and spectral tilt could easily be included. The feature values were z-standardized before training.

\subsection{Model architecture}
\begin{figure}[!t]
  \centering
  \includegraphics[width=\columnwidth]{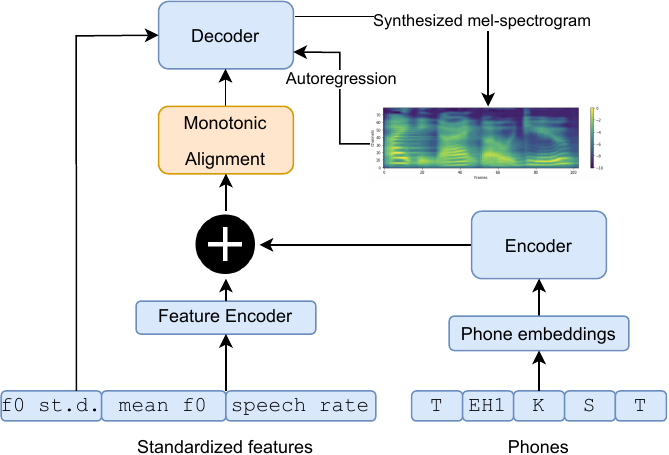}
  \caption{Model architecture}
  \label{fig:model_arch}
  \vspace{-1\baselineskip}
\end{figure}
We use a modified version of neural HMM TTS \cite{mehta2022neural}, which is an auto-regressive (AR) TTS architecture that synthesises mel-spectrograms conditioned on input text. It follows an encoder-decoder architecture similar to \cite{tacotron} but instead of a cumulative attention, it uses a left-to-right no-skip HMM (defined by a nueral network) to force monotonic alignments between text inputs and mel spectrogram frames. In addition to the already present CNN + Bi-LSTM based encoder in neural HMM we added a \emph{feature encoder} which contains a single feed-forward layer (Fig.\ \ref{fig:model_arch}) to project features into a 512-dimensional space. This modification is used to project the audio features into a more expressive control space. After standardizing the features, we use a two-step conditioning method to incorporate the prosodic features. The output of the feature encoder is first projected into the same dimensionality as phone embeddings using the feature encoder, and the two are concatenated to define the final states of the HMM. Additionally, we append a skip connection that adds the standardized prosodic features to the outputs of the encoder. The skip connection provides more robust control over the synthesis, as both encoder and decoder are conditioned on the prosodic features. We believe that this helps the model to learn the relationship between the prosodic features and the mixed representation of the phones and audio.

\subsection{Experimental setup}
To investigate the level of prosodic variation, we first performed a data analysis comparing the per-utterance mean natural logarithm of the f0 (log f0), as well as the per-utterance speech rate for each dataset.
We then trained two spontaneous TTS voice models, one baseline model trained on standard Neural HMM TTS, and our proposed model with prosodic modification, by pre-training on RyanSpeech \cite{ryanspeech} for 24,000 iterations with batch size 32, and then finetuning the model on spontaneous speech audio from the Trinity Speech and Gesture Dataset \cite{tsgd} for 9,500 iterations with batch size 20.

To examine how the control space for the prosodic features was learned, we performed an objective analysis of the synthesized utterances from our model and the baseline, in which we investigate the distribution of the features that were compared in the data analysis using gradual increments of the prosodic feature control. We synthesized 280 audio files per prosodic feature, corresponding to 40 audio files for each feature setting, modifying each standardized input feature between [-3, 3] standard deviations from the mean, while keeping the other features constant. 

We also conducted a subjective listening test in which we compared the modifiable spontaneous voice to a baseline consisting of non-modifiable spontaneous voice trained for an identical number of iterations on each corpus using the standard implementation of neural HMM TTS. 

To exemplify the system's capability to synthesize characteristic spontaneous speech with a range of voice qualities, an additional evaluation was conducted, focusing on different types of creaky phonation, which are possible to achieve implicitly by adjusting the controllable prosodic features. 

\begin{figure}[t]
    \captionsetup{justification=raggedright, width=\columnwidth}
    \includegraphics[width=0.9\columnwidth]{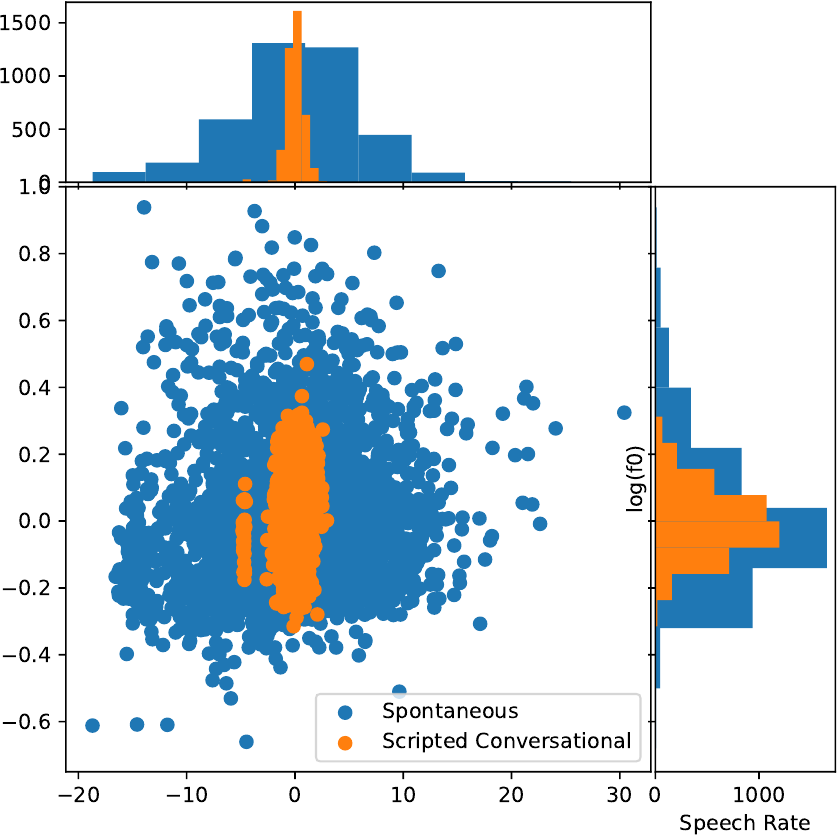}
    \caption{The mean per-utterance log f0 and mean per-utterance speech rate for the spontaneous and scripted conversational corpora}
    \label{sfig:english_dataset_stats}
    \vspace{-1\baselineskip}
\end{figure}
\section{Experiments}
\subsection{Data analysis}
To compare the distribution of the fundamental frequency and speech rate, we randomly selected 4009 audiofiles from the scripted conversational corpus to create an equal number of datapoints for each corpus. Note that the datapoints from the spontaneous corpus are measured per breath group.
Figure \ref{sfig:english_dataset_stats} shows the distribution of the mean of the per-utterance log f0, as well as the speech rate per utterance for the spontaneous and scripted conversational corpora. The values are centred around the corpus mean f0 and speech rate. In the figure, one can see that especially the speech rate is much more variable in the spontaneous speech corpus than in the scripted conversational corpus, with the speech rate for the scripted conversational corpus being closely centred around the mean, while the spontaneous corpus is a widely clustered around its mean. This is reflected in the peakier shape of the distribution for the scripted corpus.
The mean log f0 also displays more diversity and a larger range of values in the spontaneous corpus, especially for the higher-pitched datapoints. The log f0 is similarly distributed for both corpora, although it covers a larger range of values for the spontaneous corpus. 

\begin{figure*}[ht!]
   \subfloat[\label{fig:f0_std}]{%
    \includegraphics[width=0.3\textwidth, height=3.5cm]{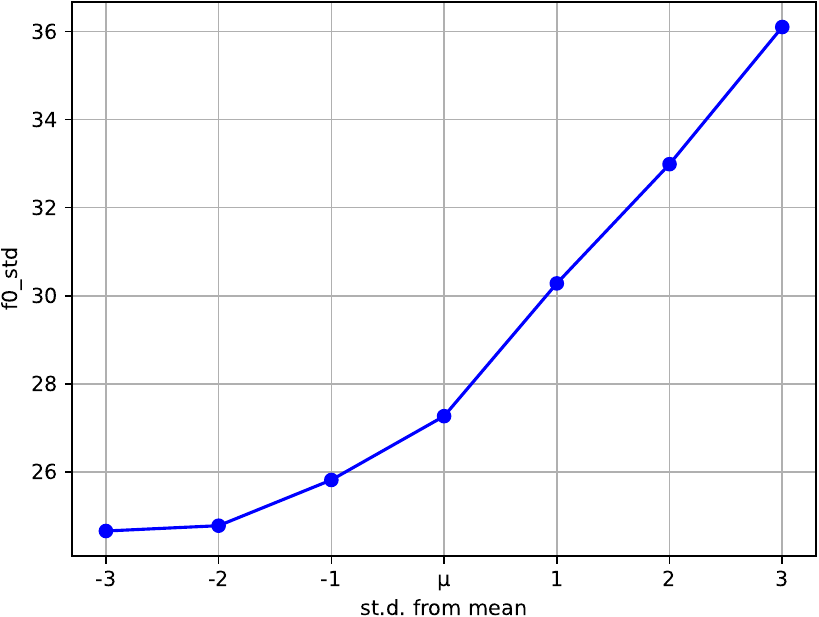}}
\hspace{\fill}
   \subfloat[\label{f0} ]{%
    \includegraphics[width=0.3\textwidth, height=3.5cm]{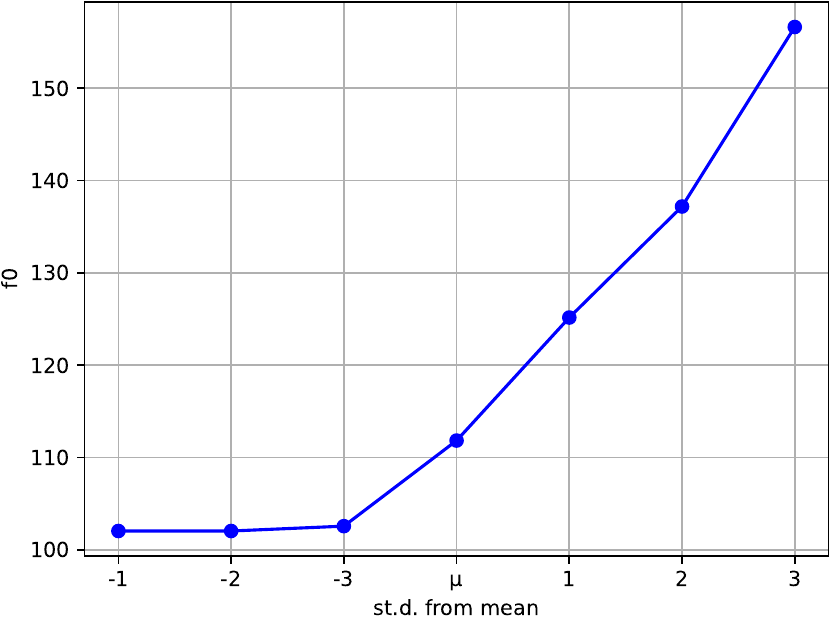}}
\hspace{\fill}
   \subfloat[\label{sr}]{%
    \includegraphics[width=0.3\textwidth, height=3.5cm]{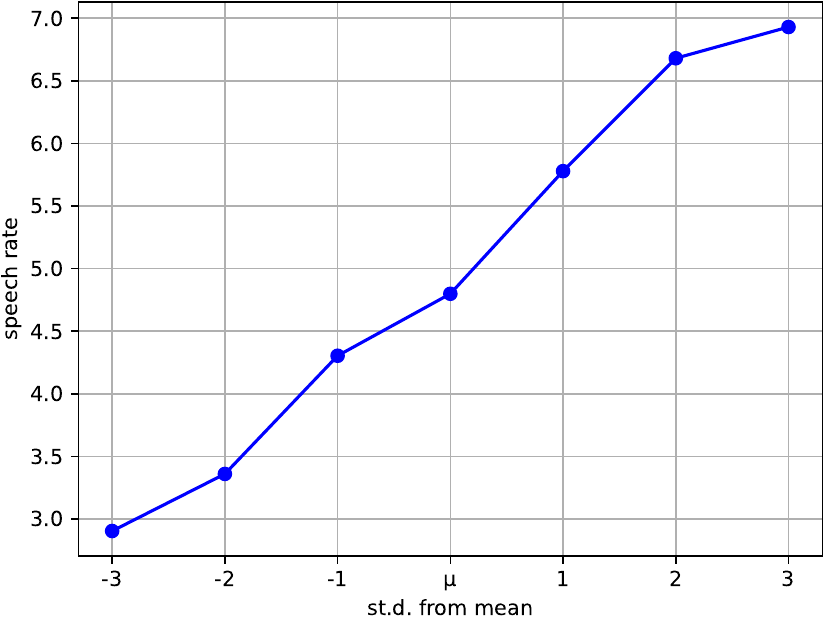}}\\
\vspace{-0.5\baselineskip}
\caption{\label{obj_eval}The effect of feature modification for (a) f0 st.d; (b) mean f0; (c) speech rate}
\vspace{-0.5\baselineskip}
\end{figure*}
\subsection{Objective analysis}
For the objective evaluation we generated 50 utterances from a held out set for various points in the feature space ranging from [-3, 3] per feature. From these utterances we computed f0 and speech rate again with WPT. The results of this evaluation are shown in Figure \ref{obj_eval}. As can be seen in \ref{fig:f0_std}, control over the f0 variability is especially predictable in the [-2, 3] standard deviations from the mean range. The smaller differences for the [-3, -2] range can be explained by the low concentration of data between (2.14\%), in addition to possibly a low occurrence of lower values in spontaneous speech.

For the mean f0, figure \ref{f0} shows there is predictable control across the [-1, 3] st.d. from the mean range. After listening to the utterances between [-3, -1] st.d. and examining the output from WPT, the cause of this is the creakiness voice quality of both the actor's and synthesized speech in this range. Creaky voice does not produce accurate f0 readings due to its lack of periodicity. We perceived -3 st.d. from the mean as having a more intense creak than -2 st.d. from the mean, suggesting that instead of lower pitch, this modification changes the level of creakiness.
Figure \ref{sr} indicates the control over the speech rate, and shows predictable control throughout the [-3, 3] st.d. from the mean range. Informal evaluation also highlighted an absence of interaction between these features, indicating the ability to vary the feature space individually for each feature. 

\subsection{Subjective evaluation}
\begin{table}
\centering
\begin{tabular}{l|r|r}
\hline
System & MOS  & Confidence Interval \\ \hline
NHMM   & 3.60 & [3.51,3.69]               \\
Proposed   & 3.48 & [3.40,3.57]               \\ \hline
\end{tabular}
\caption{ \label{tab:results} The results of the subjective MOS evaluation}
\vspace{-1\baselineskip}
\end{table}

In the subjective evaluation, the 44 native English speaking listeners recruited through Prolific were presented with 40 samples of spontaneous synthesized speech to be rated on a MOS scale. The stimuli consisted of 20 samples created with the baseline system without prosodic control, and 20 samples for which the feature values in the proposed system match the mean and variation of f0 and the speech rate of the non-modified stimulus, as extracted by WPT.
Table \ref{tab:results} shows the participants' ratings. 
The confidence intervals on the results show that the addition of prosodic control did not result in a degradation of quality.

\subsection{Evaluation of synthesis of creaky phonation}
\begin{table}
\centering
\begin{tabular}{l|r|r}
\hline
Creak Type  & Expert mean score & Measured creak \\ \hline
none   & 33.6±4.9 & 0.4\\
stylistic & 59.7±5.0 & 8.1 \\
end-of-turn & 53.8±5.0 & 7.5 \\ \hline
\end{tabular}
\caption{\label{tab:creak} The mean reported creakiness with confidence intervals and avg. creaky segments \% measured in Covarep.} 
\vspace{-1\baselineskip}
\end{table}
To evaluate the presence and naturalness of creaky voice, we synthesized 10 utterances in 3 intended styles: modal voice quality, creaky voice as a stylistic expression (throughout an utterance), and end-of-turn creak, by adjusting the features until the intended style was produced. To verify the presence of creaky phonation in the samples, we performed an analysis of the creakiness with COVAREP \cite{covarep}. In addition, we conducted a expert listening evaluation asking 15 people with training in linguistics or speech technology to rate what percentage of the utterances is creaky using a sliding scale with the range of 0-100 (Table \ref{tab:creak}). 
Participants were also asked to rate the naturalness of the creak and supply general comments. Nearly all participants rated the creakiness as natural. 
The experts commented on the influence of creakiness on the perception of the speaker's mood, mentioning that besides clear differences in the \emph{extent} of creak among the utterances, they could also distinguish between different levels of \emph{strength} within the various examples of creaky phonation.   
A one-way ANOVA shows a significant difference between these values (p$<$0.001). A post-hoc Tukey showed a significant difference between no creak, and the two styles of creak (p$<$0.001), and no significant difference between the two styles of creak (p=0.22). While the creaky utterances were indeed considered to have more creak, some utterances designed as non-creaky  were still perceived as containing some creak, possibly due to the vocal characteristics of the speaker or vocoder artefacts.    

\section{Conclusions}
We present an architecture for the prosodic modification of spontaneous speech, which is difficult to model with data-hungry attention-based architectures due to the highly complex and varied nature of spontaneous speech. We demonstrated that spontaneous speech is more varied than scripted conversational speech in terms of per-utterance mean f0, as well as speech rate. In an objective analysis we showed that our modelling of prosodic features provides control over the variation and mean of f0, and the speech rate of synthesized speech.
To show that the prosody-modifiable feature of the synthesizer does not degrade quality, we conducted a perceptual listening test in which the prosodically modified synthesized speech was rated similarly to non-modified speech.
Finally, we conducted an additional experiment to showcase the system's ability to exhaust the possibilities in varied speech data, by synthesizing natural sounding creaky voice: both as a stylistic feature, as well as in utterance-final position.

This work provides options for future research by providing an avenue for the use of more spontaneous speech corpora, which most closely correspond to real conversational speech.
The demonstrated possibility of synthesizing naturalistic and varied creaky voice lends itself to the investigation of more explicit control for voice quality dimensions.

\bibliographystyle{IEEEbib}
\bibliography{refs}

\end{document}